\documentclass[12pt,number]{elsarticle}
\usepackage{hyperref}

\pdfoutput=1 

\usepackage{color}
\usepackage[normalem]{ulem}
\usepackage{amssymb}
\usepackage{url}

\journal{Computer Physics Communications}

\begin{document}

\begin{frontmatter}
\title{Investigation of heterogeneous computing platforms for real-time data analysis in the CBM experiment}

\author{V.~Singhal\fnref{a,b}}
\ead{vikas@vecc.gov.in}
\author{S.~Chattopadhyay\fnref{a,b}}
\author{V.~Friese\fnref{c}}

\fntext[a]{Homi Bhabha National Institute, Kolkata, India}
\fntext[b]{Variable Energy Cyclotron Centre, 1/AF Bidhan Nagar, Kolkata-700 064, India}
\fntext[c]{GSI Darmstadt, D-64291 Darmstadt, Germany}

\begin{abstract}


Future experiments in high-energy physics will pose stringent requirements to computing, in particular to real-time data processing. 
As an example, the CBM experiment at FAIR Germany intends to perform online data selection exclusively in software, without using any hardware trigger, at extreme interaction rates of up to 10 MHz. 
In this article, we describe how heterogeneous computing platforms, Graphical Processing Units (GPUs) and CPUs, can be used to solve the associated computing problems on the example of the first-level event selection process sensitive to J/$\psi$ decays using muon detectors. 
We investigate and compare pure parallel computing paradigms (Posix Thread, OpenMP, MPI) and heterogeneous parallel computing paradigms (CUDA, OpenCL) on both CPU and GPU architectures and demonstrate that the problem under consideration can be accommodated with a moderate deployment of hardware resources, provided their compute power is made optimal use of.
In addition, we compare OpenCL and pure parallel computing paradigms on CPUs and show that OpenCL can be considered as a single parallel paradigm for all hardware resources.

\end{abstract}


\end{frontmatter}

\flushbottom

\section{Introduction}
\label{sec:intro}

The computing demands for modern experiments in high-energy particle and nuclear physics are increasingly challenging. 
This holds in particular for experiments relying heavily on real-time data processing.
An example is the Compressed Baryonic Matter experiment (CBM)~\cite{cbm} planned at the future Facility for Antiproton and Ion Research (FAIR)~\cite{fair} situated at Darmstadt, Germany.
CBM will study strongly interacting matter by investigating collisions of heavy nuclei at very high interaction rates of up to 10$^7$/s, enabling access to extremely rare physics probes~\cite{PhysicsPaper}.
A key feature of the experiment is a data taking concept without any hardware trigger,
which will lead to a raw data rate of about 1~TB/s~\cite{VolkerPaperOnHighRate}.
All data from the various detector systems will be forwarded to a computing farm,
where the collision data will be inspected in real-time for potentially interesting physics signatures and consequently either accepted or rejected, such that the full raw data flow is reduced to several GB/s suitable for storage.
In this article, we study the software implementation of an event selection process sensitive to events containing $J/\psi$ decay candidates. Since this process must be suitable to be deployed online, efficient
use of the available computing architectures, exploiting their parallel processing features, is mandatory~\cite{VolkerPaperOnComputationalChallange}.

In modern computing architectures, both multi-core CPU systems and auxiliary accelerators like
Graphics Processing Units (GPU) play a pivotal role. 
GPUs consist of a set of multiprocessors designed to obtain the best performance with graphics computing  
(fast mathematical calculations for the purpose of rendering images).
Nevertheless, their computational power can also be used, within certain limits, for general-purpose computing~\cite{MassivelyParallel}, an application type which is becoming more and more popular. Extracting good performance from GPU processors, however, is not trivial and requires 
architecture-specific optimizations~\cite{GpuOptimization}.

Apart from choosing the computer architecture adequate to the problem, the application developer has to select 
the programming environment making optimal use of concurrency features.
An overview and discussion of parallel programming models for both multi-core CPU systems and heterogeneous systems
is given in~\cite{ParallelProgrammingSurvey}.
On an NVIDIA GPU, the choice is between using the proprietary API CUDA, the application of which is restricted to NVIDIA hardware,
and OpenCL, an open-standard framework allowing to execute parallel code across various hardware platforms~\cite{opencl}.
Since CUDA is designed specifically for NVIDIA GPUs, one could expect that the advantage of having code portable between platforms, as offered by OpenCL, may come with a performance penalty.
Some work was thus dedicated to compare the performance of CUDA and OpenCL implementations of the same computational problem on GPU.
While e.g., Kamiri et al.~\cite{CudaOpenCLKamran} find CUDA to perform slightly better than OpenCL for their specific problem,
Fang et al.~\cite{CudaOpenCLFang} conclude, on an extensive benchmark suite, not to find worse performance for OpenCL, provided a fair comparison is done.
This situation motivates us to investigate our computing problem, the selection of $J/\psi$ candidates from heavy-ion collision events, on GPU using both CUDA and OpenCL and compare the respective performances.

Using OpenCL enables to develop portable code, which can be applied also on multi-core CPU architectures without the use
of accelerator co-processors.
This feature is of particular convenience when writing applications at a time when the architecture they will run on is not yet decided.
However, possible performance penalties with respect to native CPU concurrency standards such as OpenMP or MPI need being considered.
While some comparisons of pure parallel programming paradigms are available in the literature~\cite{MpiOpenMPPtread, MPnShPrModel},
comparisons of OpenCL to these standards are scarce; one example is to be found in~\cite{OpenCLOpenMP}.
Moreover, general conclusions are hard to be drawn; the proper choice of programming framework will depend on the nature of the concrete problem.
We thus extend our investigations by implementing the event selection algorithm in OpenMP, MPI and pthreads and confronting the performance findings with those obtained with the OpenCL implementation.

The article is organized as follows.
After a short introduction of the experiment and its computing challenge in section~\ref{sec:cbm}, 
section~\ref{sec:process} outlines the event selection algorithm.
Section~\ref{sec:computing} describes the computing platforms and testing conditions used for this study.
The implementation of the algorithm and its performance on various computing platforms and for different
parallel computing paradigms are reported in sections~\ref{sec:nvidia} and~\ref{sec:multicore}.
The results are summarized in section~\ref{sec:conclusions}, followed by an outlook and acknowledgments.

\section{The CBM experiment and its muon detection system}
\label{sec:cbm}

\begin{figure}[tbp]
	\centering
	\includegraphics[width=0.48\linewidth]{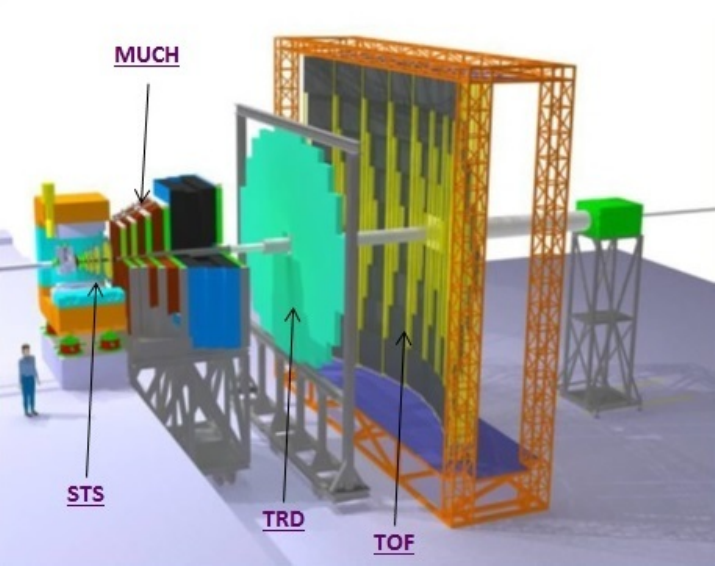}
	\hfill
	\includegraphics[width=0.48\linewidth]{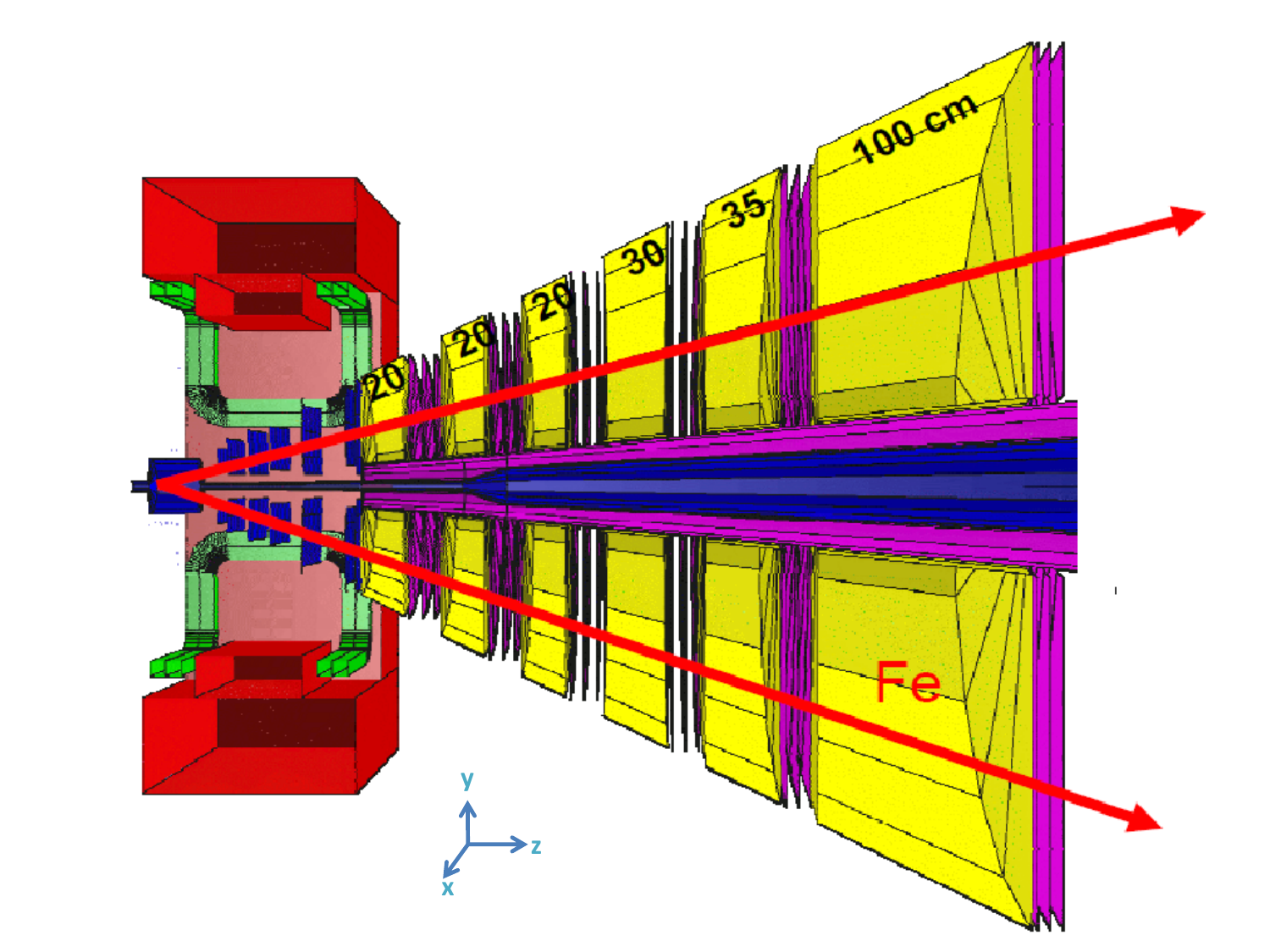}
	\caption{Left: the CBM experiment setup (muon configuration) with its detector systems; right: the CBM muon detection system with a pictorial view of a J/$\psi$ decay into $\mu^+$ and $\mu^-$. 
	Triplets of tracking stations (magenta) are placed between absorber slabs (yellow).}
	\label{faircbm}
\end{figure}

The Compressed Baryonic Matter (CBM) experiment~\cite{cbm} is a dedicated relativistic heavy-ion collision experiment at the upcoming FAIR accelerator centre at Darmstadt, Germany~\cite{fair}. For CBM, energetic ions 
($p_\mathrm{beam} = 3.5A - 35A$~GeV for Au ions) will be collided with a fixed target,
producing a large number of particles which will be registered in several detector systems serving for momentum determination and particle identification. The left panel of Fig.~\ref{faircbm} shows the experimental setup for the measurements of muons, comprising the main tracking system STS inside a magnetic dipole field, the muon detector system (MUCH), a Transition Radiation Detector used for intermediate tracking, and the Time-of-Flight Detector.

The MUCH system~\cite{muchtdr},
shown in the right-hand panel of Fig.~\ref{faircbm}, allows the separation of muons from other particles by several massive absorbers interlaid with position-sensitive detector stations.
In its full configuration, the system comprises of one graphite and five iron slabs of thicknesses varying from 20~cm to 100~cm.
The absorbers are separated by a 30 cm gap housing three layers of position-sensitive gas detectors. The distance between two detector layers is 10 cm. The last station consisting of a detector triplet is called trigger station and is positioned after a one meter thick iron absorber.  
The detectors layers are constructed from GEM modules with a pad segmentation in $r-\phi$ geometry. The pad size varies in the radial direction according to the hit density profile, such as to obtain an approximately uniform occupancy. The full system comprises about $5 \cdot 10^5$ readout channels. It is currently investigated to use RPC instead of GEM detectors in the two downstream stations for costs reasons; in this study, GEM detectors are assumed throughout. The position resolution depends on the radial distance from the beam line with corresponding pad size e.g. station-1 pad size varies from 0.32cm to 1.71cm.

One objective of the experiment is the measurement of J/$\psi$ meson production via their decay into a muon pair ($\mathrm{J}/\psi \to \mu^+ \mu^-$).  At CBM energies, however, the J/$\psi$ multiplicity is expected to be extremely small - of the order of $10^{-7}$ per collision~\cite{ParthaPaper}.  The experiment will thus be operated at extreme interaction rates in order to obtain sufficient statistics. To reduce the total data rate of about 1 TB/s to a recordable value of a few GB/s, J/$\psi$ candidate events must be selected in real-time during data taking, rejecting the vast majority of the collisions.
The J/$\psi$ trigger signature is obvious from the right panel of Fig.~\ref{faircbm}: two muons traversing the absorber system and being registered simultaneously in the trigger station. The trajectories of the muon candidates have to point back to the collision vertex (target), since the J/$\psi$ decays promptly at the vertex \cite{ParthaThesis}.

The trigger signature will be evaluated exclusively in software on the so-called FLES (First-Level Event Selection) computing cluster~\cite{fles, fleslink} hosted in the Green Cube building at GSI. 
Efficient trigger algorithms are a prerequisite for the affordability of the FLES cluster. 

\section{The real-time event selection process}
\label{sec:process}

\subsection{Trigger signature}

The signature for J/$\psi \to \mu^+ \mu^-$ candidate events is a rather simple one and is visualized in the right panel of Fig.~\ref{faircbm}.
The two daughter muons, having high momentum because of the large $q$ value of the decay, traverse all absorber layers and reach the trigger station, while hadrons, electrons, and low-momentum muons will be absorbed. 
Since J/$\psi$ decays promptly ($\mathrm{c}\tau = 7.1 \cdot 10^{-21} \mathrm{s}$), the decay products practically originate from the primary (collision) vertex, i.e., from the target.
Owing again to the high momentum of the muons, their trajectories can be approximated by straight lines even in the bending plane of the dipole magnetic field, which has a bending power of 1~Tm \cite{magnettdr}. The trigger station consisting of three detector layers provides three position measurements, allowing to check the back-pointing to the primary vertex.
The signature of a candidate event is thus the simultaneous registration of two particles in the trigger station which can be extrapolated backward to the target.

\begin{figure}[tbp]
	\centering
	\includegraphics[height=0.39\linewidth]{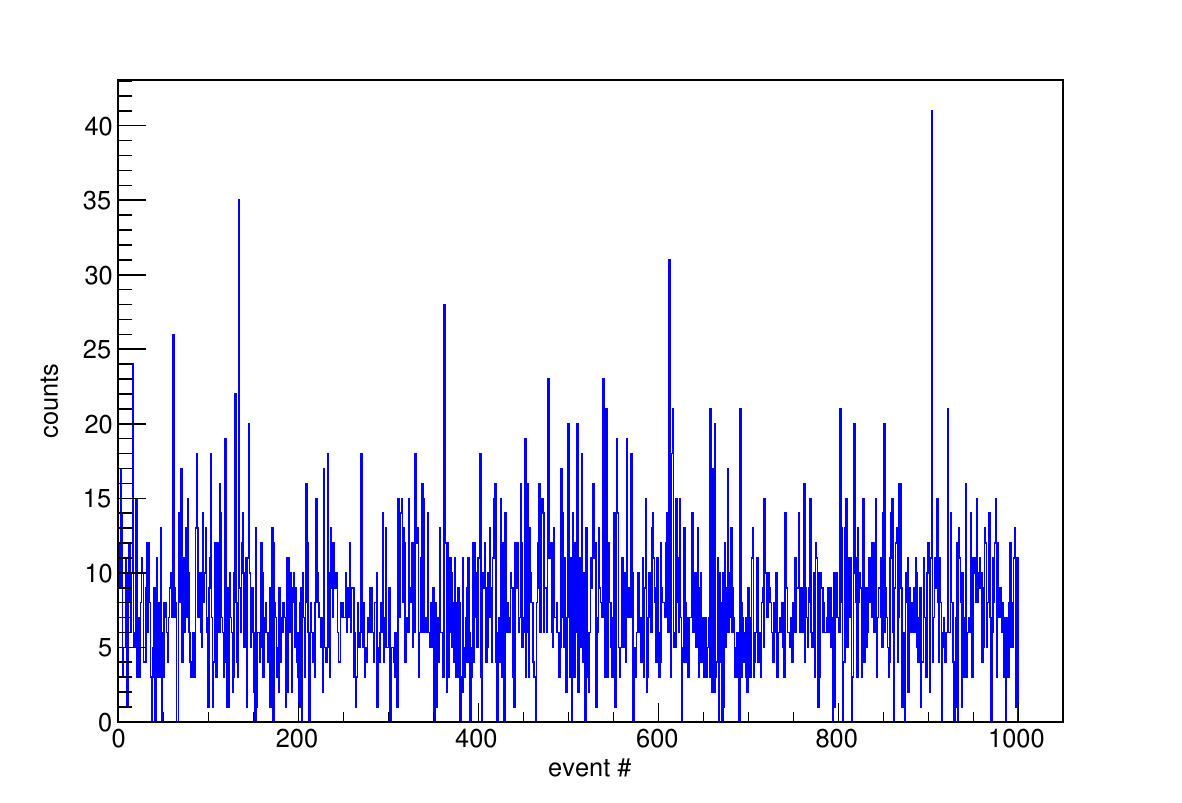}
	\hfill
	\includegraphics[height=0.39\linewidth]{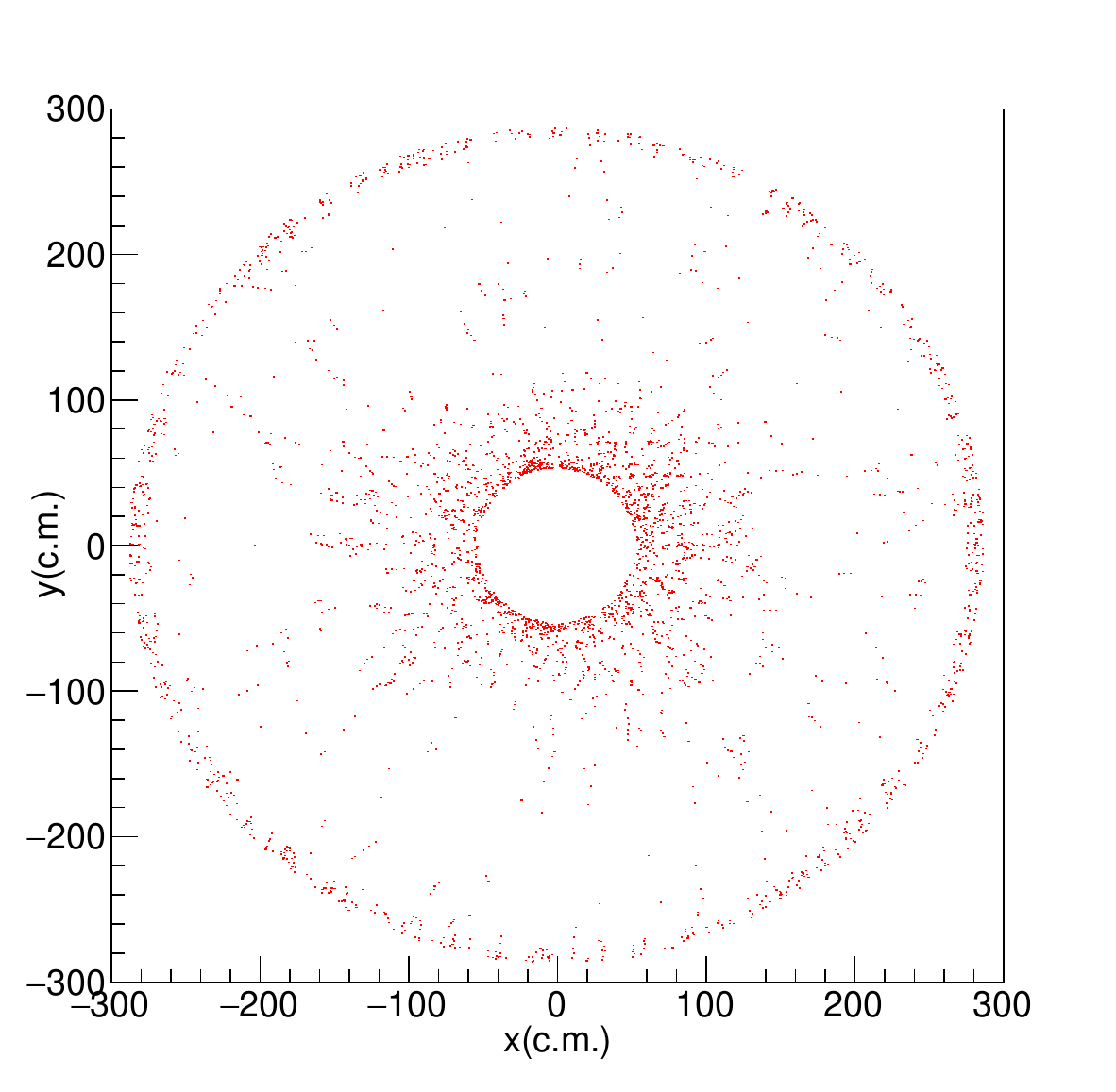}
	\caption{Left: event-wise hit distribution on the last MUCH station (combined of the last three layers); right: $x-y$ distribution of background hits in the trigger station for 1000 events. The accumulation at the periphery is due to secondaries not  shielded by the absorber.
	}
	\label{HistogramAndDistribution}
\end{figure}

To study the feasibility of this selection procedure, Fig.~\ref{HistogramAndDistribution} (left) shows the number of hits per event for the trigger station (combined for all three layers) for a sample of background events (not containing any J/$\psi$ decay) simulated in the CBM setup.
The average number is about 45 for the trigger station and 15 for each of its layers. 
The dominant sources of this background are secondary muons originating from weak decays of $\Lambda$ and $K^0_\mathrm{S}$ between the target and the trigger station. 
The right panel of Fig.~\ref{HistogramAndDistribution} shows the spatial distribution in the transverse plane ($x-y$) of these background hits.
The void area in the centre is not instrumented, since it hosts the vacuum pipe for the non-interacting beam \cite{muchtdr}.

\subsection{Algorithm}

\begin{figure}[tbp]
	\centering
	\includegraphics[width=0.48\linewidth]{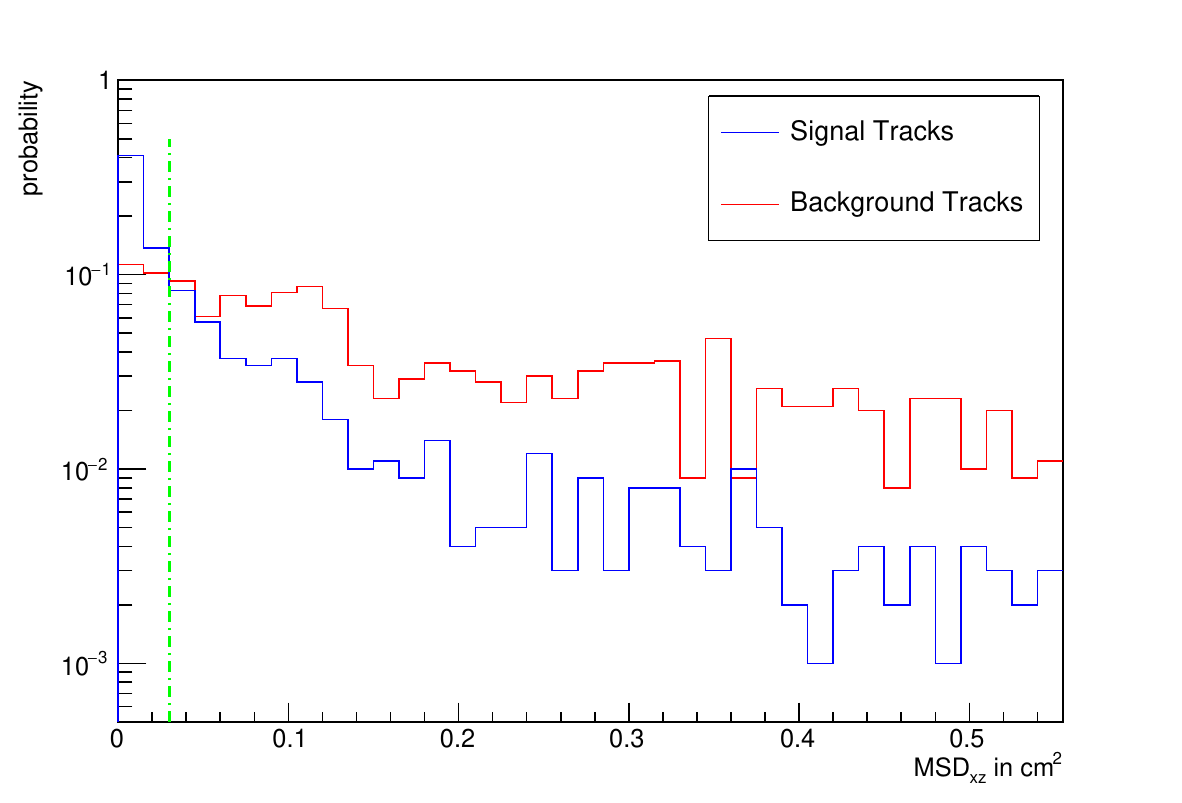}
	\hfill
	\includegraphics[width=0.48\linewidth]{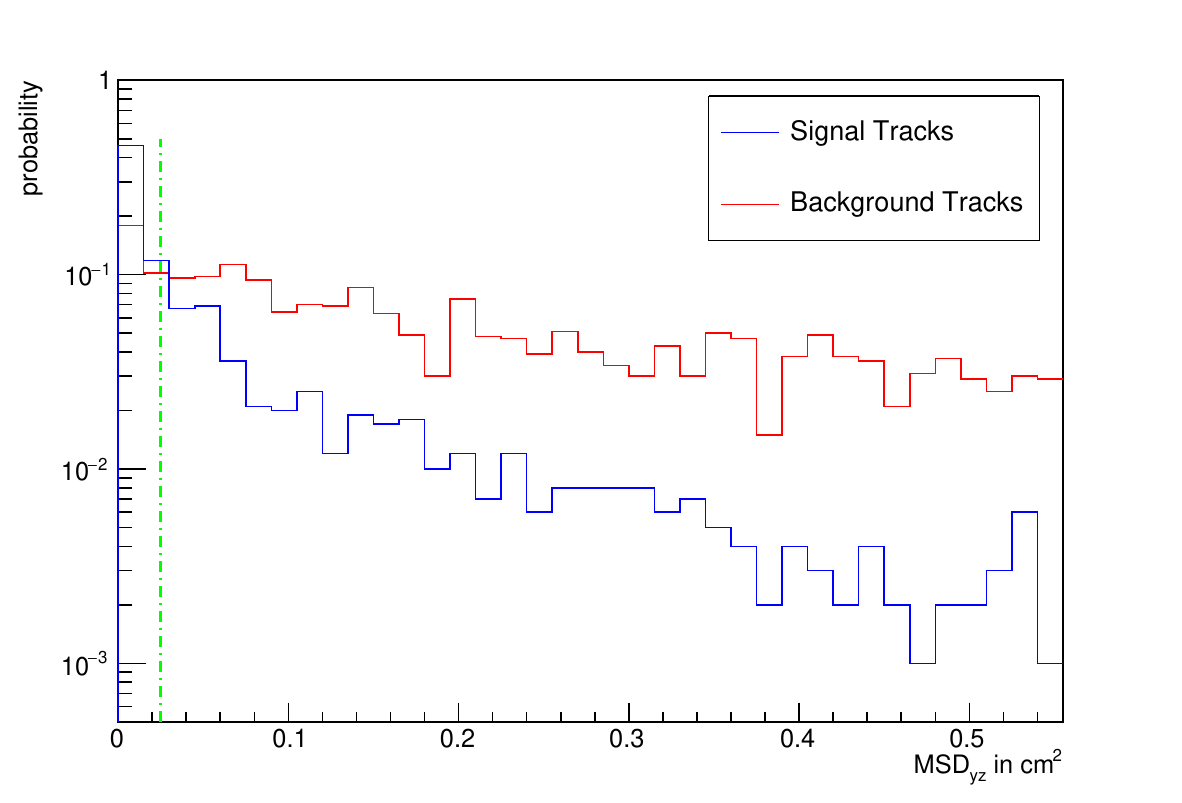}
	\caption{ Left: ($MSD_{xz}$) and right: ($MSD_{yz}$) event normalised distribution for signal tracks in 1000 signal events (Au+Au at 35$A$~GeV) overlay with all triplets in background events with vertical (green) cut line.}
	\label{XChiSquareDistribution}
\end{figure}

The algorithm corresponding to the signature described in the previous section operates on all
data from the MUCH detector within one event (collision). It comprises the following steps:

\begin{enumerate}
\item Create all possible triplets of hits in the trigger station, with one hit from each layer.
\item For each triplet:
\begin{enumerate}
\item Fit the hits of the triplet plus the event vertex (0, 0) by a straight line in both the $x-z$ (bending) plane and the $y-z$ (non-bending) plane, i.e., $x=m_{0} z$ and $y= m_{1} z$.
\item Compute the Mean Square Deviation (MSD) of the triplet fit. A triplet is rejected if the MSD (per degree of freedom) is above a threshold (0.03 for $MSD_\mathrm{xz}$, 0.025 for $MSD_\mathrm{yz}$). 
The MSD cut suppresses random combinations of hits as well as real triplets from secondary tracks.
To illustrate, Fig.~\ref{XChiSquareDistribution} shows the MSD distribution for the $x-z$ plane (left) and the $y-z$ plane (right) for signal tracks (central Au+Au events at $p_\mathrm{beam} = 35A$~GeV)(blue) with overlay of all triplets in background events (red) and a vertical (green) line indicate the cut value.

The threshold values result from an optimization procedure, taking into account signal efficiency and background suppression.
\end{enumerate}
\item Select the event if it contains at least two triplets passing the MSD cut.
\end{enumerate}

For the application to simulated events used in the following, data are read from a file produced by the detector simulation (GEANT3~\cite{geant3} or GEANT4~\cite{geant4} plus detailed detector response implementation). 
When deployed for data taking, the algorithm will receive the input (hit) data from the online data stream, which is aggregated by the data acquisition software and supplied to the compute nodes through remote direct memory access (RDMA)~\cite{LamannaPaper}.
Hit data are grouped into events; they provide three-dimensional coordinate information ($x, y, z$).

\subsection{Evaluation of the trigger algorithm}

In order to assess the performance of the trigger algorithm, the following sets of simulated data were produced and studied:
\begin{enumerate}
\item[D1.] Signal events containing only one decay J/$\psi \to \mu^+ \mu^-$. The phase-space distribution of the J/$\psi$ were generated using the PLUTO generator~\cite{pluto}.
\item[D2.] Background events (central Au+Au at $p_\mathrm{beam} = 35A$~GeV) generated using the UrQMD model~\cite{urqmd}.
\item[D3.] Background events with one embedded decay J/$\psi \to \mu^+ \mu^-$ in every UrQMD event.
\end{enumerate}
The following performance figures are used:
\begin{enumerate}
\item[a)] efficiency (E): the fraction of embedded signal events selected by the algorithm;
\item[b)] efficiency under acceptance (EUA): the fraction of selected embedded signal events in which both decay muons have hits in all three layers of the trigger station;
\item[c)] background suppression factor (BSF): the ratio of all background events to background events selected by the trigger algorithm. 
\end{enumerate}
The efficiency is mainly determined by the geometrical coverage of the muon detection system. Typical values are about 39~\%.
With the cut values named above, the EUA is 85.4~\%.
The reason for it to be smaller than unity is one or both signal tracks not passing the MSD cut.
The BSF of 71.4 shows that the primary aim of the algorithm - suppression of a large fraction of the input data rate - is reached; the probability to find a chance pair of triplets passing the MSD cut is still not negligible. 

Further background suppression will be achieved by full track reconstruction in the STS and MUCH detectors.
This will provide the track momentum and more precise determination of the track
impact parameter on the target plane, allowing to better separate primary from secondary muons.
Full track reconstruction, however, is algorithmically involved~\cite{ca} and thus requires significant
computing resources. Owing to the first-level trigger algorithm described here, it needs only be applied on a data rate already reduced by a factor of about 70.

\section{Heterogeneous computing}
\label{sec:computing}

Modern computers come with a variety of concepts for concurrent data processing on many-core architectures~\cite{Multicore}: dedicated co-processors like NVIDIA or AMD GPUs, Many Integrated Core (MIC) by Intel such as Xeon Phi, the Accelerated Processing Unit (APU) of AMD, the Cell Processor of IBM and others. Nowadays, a single system comprising more than one host CPU and more than one GPU is called a heterogeneous system~\cite{GpuEvolution}. The architectures of such systems  are based on the SIMT (Single Instruction Multiple Thread) technology and come with a plenitude of processing units, allowing to run many threads in parallel. 
However, in order to make use of their computing potential, adequate programming paradigms are needed.
The pure parallel programming paradigms Posix Threads (pthread)~\cite{PosixStandard}, OpenMP~\cite{openmp} and MPI (Massage Passing Interface)~\cite{mpi,mpi2extn,mpi2} are available since long for utilizing multi-core CPU architectures like those of Intel, AMD, or IBM. 
For many-core architectures, Apple developed the OpenCL (Open Compute Language) managed by the Khronos group~\cite{opencl}, and NVIDIA came with CUDA~\cite{cuda,cudabook}, a proprietary parallel programming API that can be used for NVIDIA hardware only.

The choice of architecture and programming paradigm may well depend on the specific computing problem to be solved.
For the algorithm described in the previous section, we have tested implementations on the following two heterogeneous platforms:
\begin{enumerate}
\item[S1.] A Dell T7500 workstation comprising two Intel Xeon 2.8~GHz six-core processors with 2~GB/core RAM, together with two NVIDIA GPUs (Tesla C2075~\cite{c2075} and Quadro 4000~\cite{quadro}).
\item[S2.] An AMD-based HP Server with four AMD Opteron 2.6~GHz processors comprising 16 cores with 4~GB/core RAM (total 64 cores). 
\end{enumerate}
The algorithm was implemented on these setups with different pure and heterogeneous parallel programming models, namely:
\begin{enumerate}
\item[a)] Using both Intel processors (12 cores in total) of the S1 setup with pthread, OpenMP, MPI and OpenCL;
\item[b)] Using all four AMD processors (64 cores in total) of the S2 setup with pthread, OpenMP, MPI and OpenCL;
\item[c)] Using the Tesla C2075 GPU of the S1 setup (448 cores) with CUDA and OpenCL;
\item[d)] Using the Quadro 4000 GPU of the S1 setup (256 cores) with CUDA and OpenCL.
\end{enumerate}

The event selection process described in section~\ref{sec:process} was implemented using OpenCL~1.1 on both CPU and GPU architectures, and CUDA 4.0 on the NVIDIA GPUs. 
In addition, it was implemented on CPUs with pthread, OpenMP~3.1 and MPI~3.3. 
The results presented in this study were obtained with Scientific Linux CERN release 6.10 (Carbon), kernel version 2.6.32-754.el6.x86\_64, on the CPUs. We used the gcc compiler 4.8.2 with the optimization option \textit{-O2} 
 and the architecture-specific option \textit{-march=native}, which we found to provide the best results~\cite{AarushiPaper}.
Timing measurements were performed using \textit{std::chrono} library functions. 
As relevant performance figures, we
report average execution time per event or the average throughput. 
Note that the thread creation, distribution and aggregation times are not accounted for in MPI timing measurements whereas they are for pthread, OpenMP and OpenCL on CPU.

\section{Investigation on NVIDIA GPUs}
\label{sec:nvidia}
In this section, we explore the NVIDIA GPU architecture, memory arrangements and the implementation and optimization of the event selection process on NVIDIA GPUs. We compare different heterogeneous parallel programming paradigms (CUDA and OpenCL) on NVIDIA's Tesla and Quadro GPUs.  

\subsection{NVIDIA GPU C2075 and thread architecture}

\begin{figure}[tbp]
	\centering
	\includegraphics[scale=0.45]{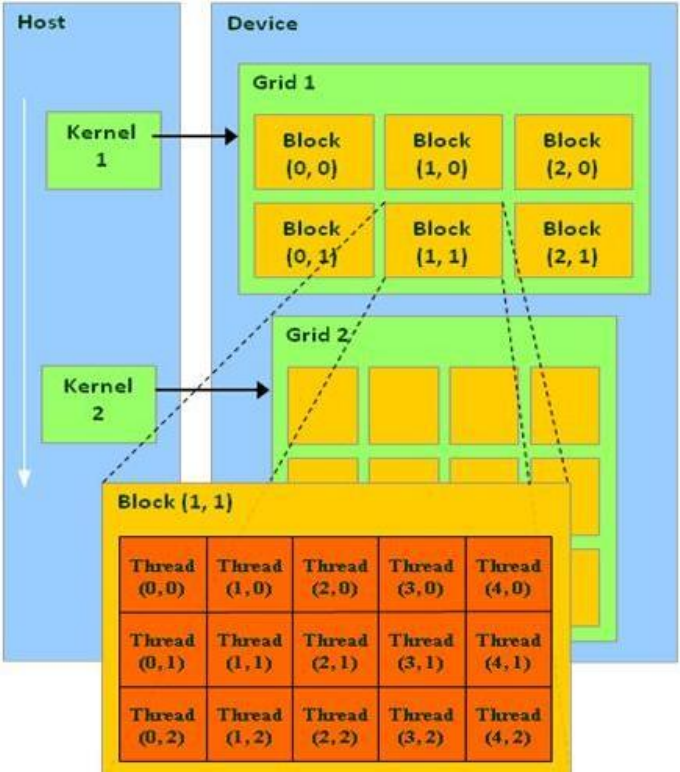}
	\caption{Grid, block and thread architecture of a GPU~\cite{gridthreadimage}}
	\label{threadarchitecture}
\end{figure}

The Tesla C2075 GPU~\cite{c2075} used in this work consists of an array of 14 so-called Streaming Multiprocessors (SM). Each SM contains 32 Scalar Processor (SP) cores. The smallest executable unit of parallelism on a GPU comprises of 32 threads known as a warp of threads.  
Therefore, in total $14 \cdot 32 = 448$ cores  are available in a single GPU card. 
All SPs within an SM share resources such as registers and memory. The Instruction Issue Unit distributes the same instruction to each SP inside a SM. 
Thus, they execute the same instruction at any time, a concept known as SIMT (Single Instruction Multiple Threads). 

Figure~\ref{threadarchitecture} shows a diagrammatic representation of the GPU architecture.
The thread hierarchy is of two levels. 
At the topmost level there exists a grid of thread blocks. 
At the second level, the thread blocks are organized as an array of threads. 
A function to be executed on the GPUs is known as kernel. 
Kernel execution takes place in the form of a batch of threads organized as a grid of thread blocks. 
The thread blocks are scheduled across SMs. 
Each block comprises of many warps of 32 threads. 
Threads belonging to the same warp execute the same instruction over different data.
The efficiency of computation is best when the threads follow the same execution path for the majority of the computation~\cite{GpuBenchmarking}. 
Execution divergence, when threads of a warp follow different execution paths, is handled automatically inside the hardware with a slight penalty on execution time. 
The size of thread blocks (number of threads per block) and number of blocks can be managed by the programmer.

\subsection{The CUDA platform and memory hierarchy for the Tesla C2075 GPU}

The Compute Unified Device Architecture (CUDA) has been developed by NVIDIA for performing general-purpose computing on NVIDIA GPU using parallel computation features. The CUDA memory organization is hierarchical. 
Each thread has its own private local memory apart from the registers, which are 32,768 per SM.
The Tesla C2075 GPU includes a configurable (as instruction or data) L1 cache per SM block and a unified L2 cache for all processor cores.  
All threads inside a block share different memory spaces - per block shared memory. 
The size of a block local shared memory is 48~KB. 
Its lifetime equals that of the block and is characterized by low memory access times. 
Shared memory comprises of a sequence of 32-bit words called banks. 
There also exists a global memory (6 GB) shared by all threads across all thread blocks, having the lifetime of the application. 
The access time to the global memory is larger than that to other memories. 
Global memory comprises of 128 byte segment sequence, and at any time, memory requests for 16 threads (a half warp) are serviced together. 
Each segment corresponds to a memory transaction. 
If the threads in a half warp access data spread across different memory segments (uncoalesced memory request), the corresponding multiple memory transactions would lower the performance~\cite{MemoryCoalescing}.

\subsection{Investigation with CUDA}
\label{subsec:cuda}

The event selection algorithm described in section~\ref{sec:process} was implemented in the C language using the CUDA API and then compiled with the NVIDIA compiler (nvcc v7.0.27).
To optimize the event selection code for the GPU, we analysed the program and performed memory arrangement according to the GPU architecture as discussed above. Multiple events are processed at the same time, one event being allocated to one thread. After development and implementation of the event selection process using CUDA, we concentrated on optimizing the CPU to GPU data transfer time, which is significant as the data volume is large. 

In our first approach, the entire event hit data of the setup, consisting of the STS and MUCH detectors, was transferred to the GPU.
The following steps were taken:
\begin{enumerate}
\item Data are read in from a file; in the actual experiment, the data will be deployed in shared memory by the data acquisition software~\cite{fles}. The file I/O times are thus not included in the following timing measurements.
Since access to the shared memory will be performed in parallel to the algorithmic computation, we do not expect a significant impact on the final result.

\item Memory on the GPU device is allocated for a chosen number of events $n_\mathrm{ev}$(cudaMalloc).
			
\item Hit data for $n_\mathrm{ev}$ events are transferred from CPU to GPU (cudaMemcpy).

\item The number of blocks $b$ and the number of threads per block $t$ are selected to optimise the GPU computation time.

\item The event selection algorithm is executed in GPU threads for $n_\mathrm{ev}$ events in parallel, with $n_\mathrm{ev}  \leq b \cdot t$.

\item The list of selected events is transferred back to the CPU (cudaMemcpy).

\end{enumerate}

The GPU schedules and balances the selected number of blocks and threads on the available SMs (14 for C2075) and SPs (32).
For processing, the data container with the hit coordinate array is arranged as \\
{
$x_{11}$, $y_{11}$, $z_{11}$, $x_{12}$, $y_{12}$, $z_{12}$, ....... $x_{1n}$, $y_{1n}$, $z_{1n}$  \\
$x_{21}$, $y_{21}$, $z_{21}$, $x_{22}$, $y_{22}$, $z_{22}$, ....... $x_{2n}$, $y_{2n}$, $z_{2n}$ \\
........ \\
$x_{m1}$, $y_{m1}$, $z_{m1}$, $x_{m2}$, $y_{m2}$, $z_{m2}$, ....... $x_{mn}$, $y_{mn}$, $z_{mn}$
} \\
where $x, y, z$ are the hit coordinates in configuration space; the first index denotes the event, the second one the consecutive number of the hit in the event.
 
Our investigation showed that this data arrangement suffers from uncoalesced memory access to the $x, y, z$ coordinates (see also~\cite{MemoryCoalescing}). 
By its SIMT architecture, CUDA executes 32 threads of a block simultaneously; 
therefore all 32 threads should read from the global memory in a single or double read instruction. 
To cope with this, we rearranged the data such that coalesced memory access is possible. 
First, we introduced separate data containers for each coordinate axis, and second, hits of different events are arranged together: \\
{
$x_{11}$, $x_{21}$,........$x_{m1}$, $x_{12}$, $x_{22}$, ........$x_{m2}$................$x_{1n}$, $x_{2n}$........$x_{mn}$ \\
$y_{11}$, $y_{21}$,........$y_{m1}$, $y_{12}$, $y_{22}$, ........$y_{m2}$................$y_{1n}$, $y_{2n}$........$y_{mn}$ \\
$z_{11}$, $z_{21}$,........$z_{m1}$, $z_{12}$, $z_{22}$, ........$z_{m2}$................$z_{1n}$, $z_{2n}$........$z_{mn}$ \\
}

In the course of further optimizing the process, we found that the 
majority of time is taken by the global read of data by each thread, thereby requiring a reduction in global read time as data reside in the global memory and not in the shared or private memory of the GPU. 
By construction of the algorithm, the number of global reads for each thread is proportional to the number of events $n_\mathrm{ev}$. 
Each event contains about 5000 hits, and every global read takes around 300--400 clock cycles~\cite{GpuBenchmarking}. 
For the computation, however, only a small fraction of these data are used, namely hits in the last (trigger) station, which are about 15 per layer per event (see Fig.~\ref{HistogramAndDistribution}). 
Thus, we introduced a filtering of the data on the CPU host side, such that only hit data in the trigger station are transferred to GPU~\cite{vikas}.

\begin{figure}[tbp]
	\centering	
	\includegraphics[width=0.48\linewidth]{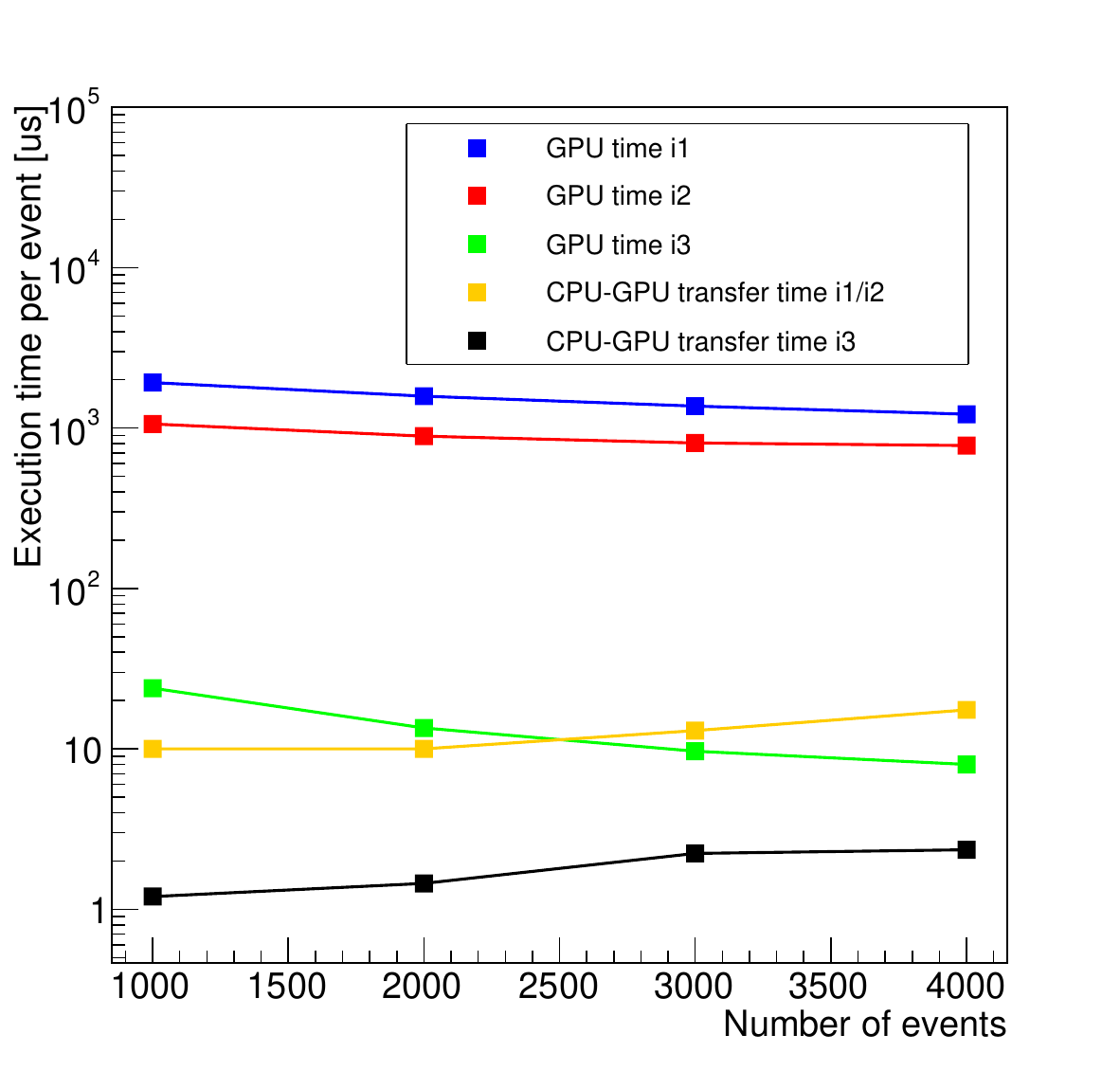}
	\hfill
	\includegraphics[width=0.48\linewidth]{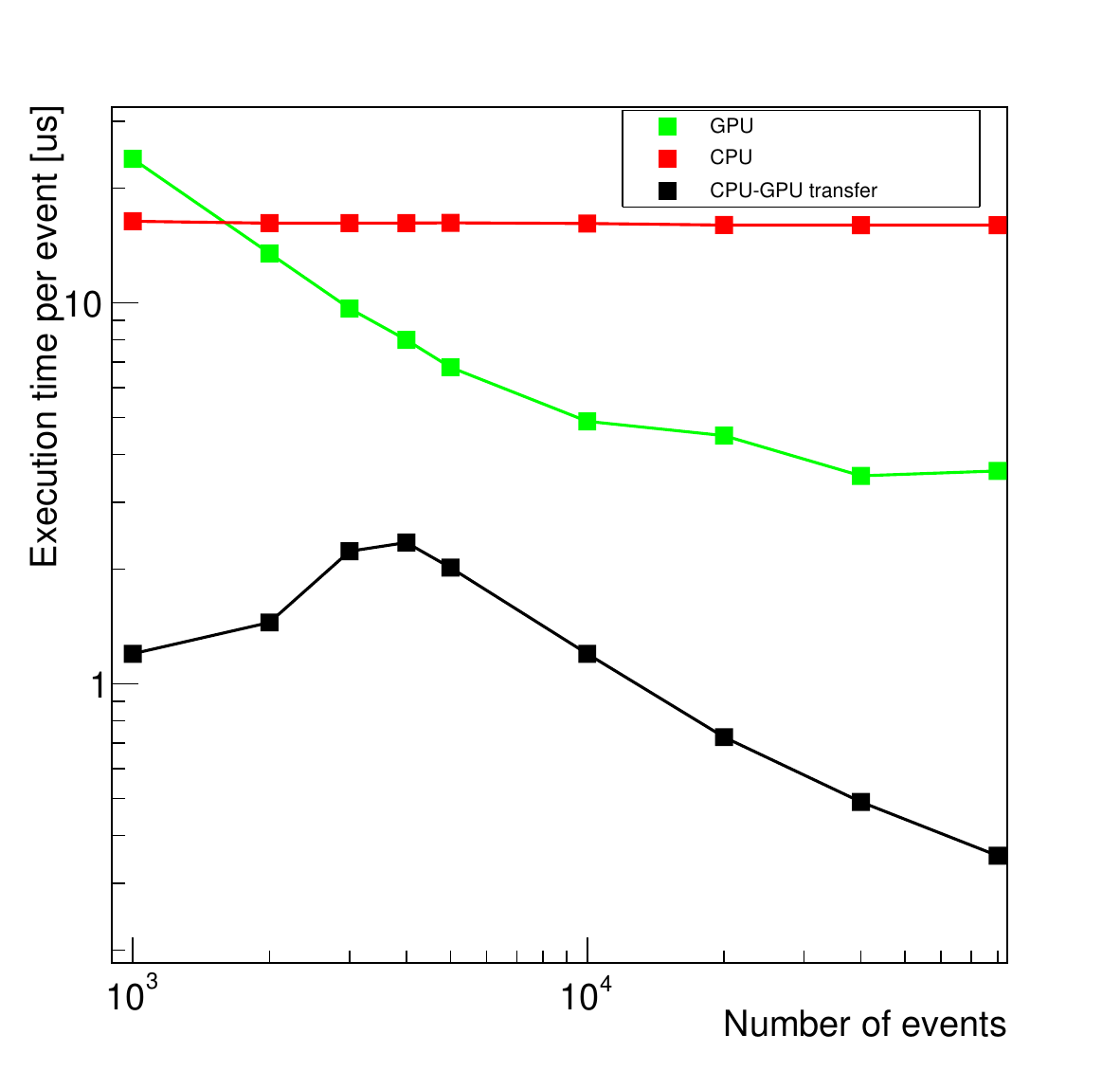}
	\caption{Processing time per event in microseconds as function of the number of events. The left panel compares the implementations i1, i2 and i3 with CUDA on the Tesla GPU (see text). 
	A comparison of the execution times on GPU (implementation i3) and on CPU (single-thread) is shown in the right panel.}
	\label{compare}
\end{figure}

The importance of the optimization steps is illustrated in Fig.~\ref{compare} (left), showing the per-event GPU execution time for the various implementations on the Tesla GPU and respectively the 
per-event CPU to GPU data transfer time. 
This study was performed for up to 4,000 events because of memory limitations of the GPU. 
The processing time is reduced by a factor of two
from the first implementation (i1) to the one properly using coalesced memory (i2).
The data transfer time is the same for both implementations
since the same data are transferred. 
Filtering of input data at the host side (i3) gives a reduction by about two orders of magnitude for the per event-execution time compared to (i2) and one order of magnitude for the per-event data transfer time.

\begin{table}[h]
	\begin{center}
	\caption{\label{tab:resultsGPU}Results for the event selection algorithm on the Tesla GPU}
	\begin{tabular*}{5.0in}{@{\extracolsep\fill}p{.45in}p{.45in}p{.45in}p{.5in}p{.7in}p{.5in}p{1in}}
	\hline
	\hline
	\# Events & \# blocks & \# threads & GPU Time (ms)& CPU-GPU Transfer Time(ms)& CPU Time (ms)& Speed-Up (CPU time/GPU time) \\ \hline
	1000 & 32 & 32 & 23.9 & 1.2 & 16.38 & 0.69\\
	2000 & 64 & 32 & 27 & 2.9 & 32.40 & 1.20\\
	3000 & 64 & 64 & 29 & 6.7 & 48.60 & 1.68\\
	4000 & 64 & 64 & 32 & 9.4 & 64.83 & 2.03\\
	5000 & 128 & 64 & 33.9 & 10.1 & 81.16 & 2.39\\
	10000 & 128 & 128 & 48.9 & 12 & 161.67 & 3.31\\
	20000 & 256 & 128 & 89.7 & 14.5 & 320.24 & 3.57\\
	40000 & 512 & 128 & 140.7 & 19.7 & 640.25 & 4.55\\
	80000 & 1024 & 128 & 289.8 & 28.3 & 1280.39 & 4.42\\
	\hline
	\hline
	\end{tabular*}
	\end{center}
\end{table}

Figure~\ref{compare} (right) and Table~\ref{tab:resultsGPU} compare the per-event GPU execution time and data transfer time for implementation i3 to the single-threaded execution time on CPU.
The data filtering on the host side relaxes the restrictions imposed by the limited GPU memory, such that a larger number
of events (we tested up to 80,000) can be processed at a time.
The data transfer time is lower by one order of magnitude compared to the execution time; 
moreover, it can be hidden by performing computation and transfer in parallel~\cite{LamannaPaper}.
The measurements demonstrate the importance to load the GPU with sufficient data in order to make optimal use of its capacity.
Compared to the single-threaded execution on the CPU (using optimization in the gcc compiler), we obtain a speed-up of 4.55 for a data set of 40k events by using the Tesla GPU. Table~\ref{tab:resultsGPU} shows that for more than 40k events, the speed-up with respect to the single-threaded CPU is slightly reduced,
indicating that the optimal data load on the GPU is reached with this amount of events. 
Our investigations show that about $3 \cdot 10^5$ events per second can be processed on a single Tesla GPU.

\subsection{Implementation with OpenCL and comparison with CUDA}

The previous section has demonstrated that making optimal use of a GPU with the CUDA API is far from trivial
and requires sophisticated optimization of the data arrangement. 
The OpenCL programming paradigm~\cite{opencl} offers an architecture-independent alternative.
However, for a beginner OpenCL seems difficult as far as its syntax and programming procedure are concerned. 
Writing a small ``Hello World'' program in OpenCL needs creating platform, device, context, and command queue, then memory allocation via create and writing buffer, program object creation via creating source and building program, program execution via create kernel, enqueueing kernel, reading back buffer, etc. 
At first sight, this seems cumbersome compared to CUDA which provides an easy terminology for writing programs~\cite{cudabook}. 
On the other hand, OpenCL programs can be compiled via available C or C++ compilers, unlike CUDA which requires a vendor-specific compiler. 

\begin{figure}[tbp]
	\begin{center}
	\includegraphics[width=0.6\linewidth]{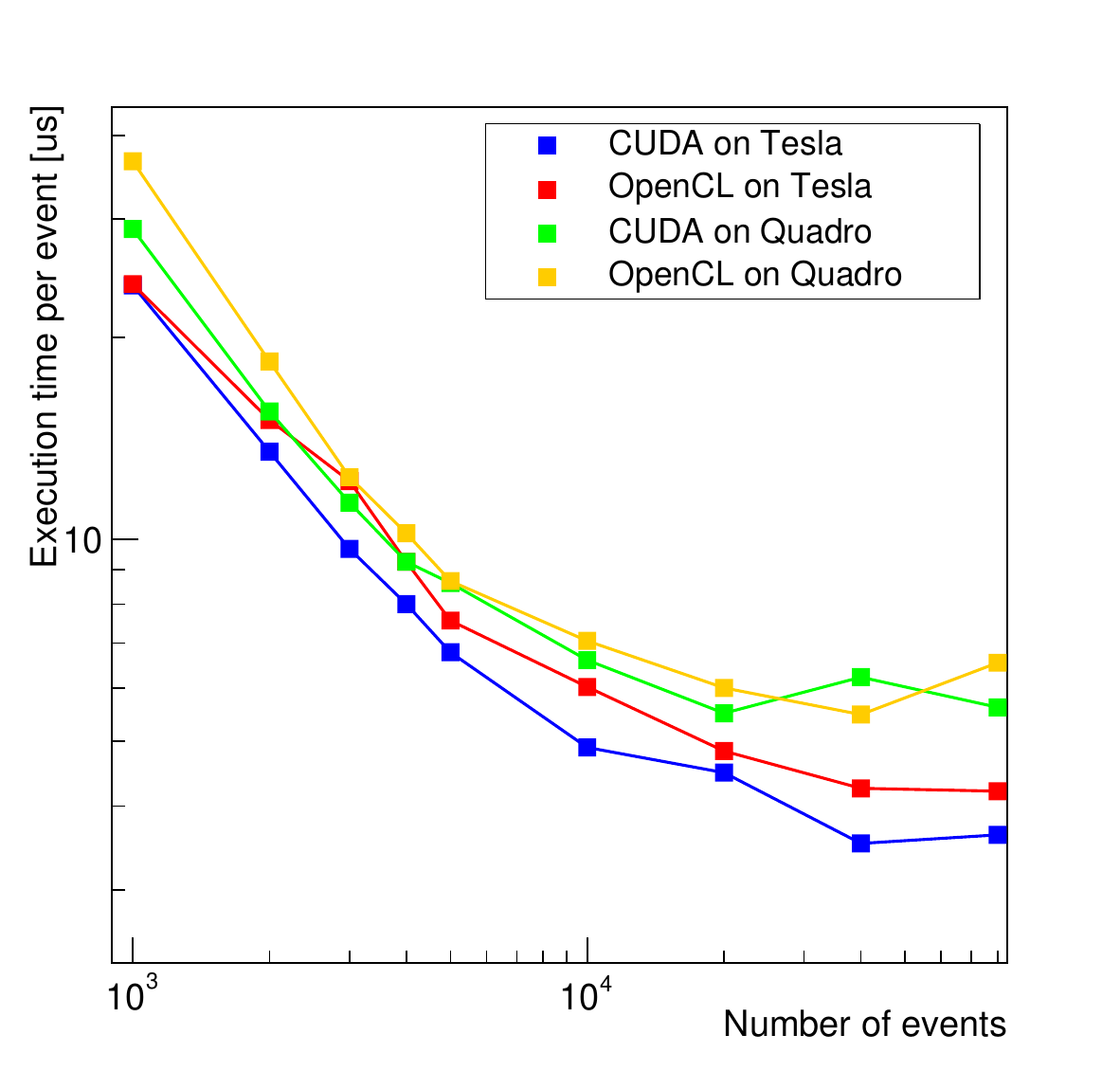}
	\caption{Execution time per event for the CUDA and OpenCL implementations on the NVIDIA Tesla and Quadro GPUs}	
	\label{CompareCUDAOpenCL}
	\end{center}
\end{figure}

Once OpenCL programs are written and compiled, then can be executed on any device, whether GPU, CPU, or APU, whereas CUDA can be executed only on NVIDIA GPUs. 
Both CUDA and OpenCL treat the CPU as host, but for CUDA only the NVIDIA GPU is a device, 
whereas OpenCL treats any hardware as computing device by creating an instruction queue that can be executed on all available computing resources. We thus investigated OpenCL as an open-source solution for heterogeneous programming
in the spirit of the studies presented in~\cite{CudaOpenCLKamran} and~\cite{CudaOpenCLFang}.

Figure~\ref{CompareCUDAOpenCL} shows the per-event execution time of the event selection algorithm for different numbers of events on the Tesla and Quadro GPUs using the implementations in CUDA and in OpenCL. 
We find the OpenCL code execution time to be slightly higher than that of the CUDA code on both Tesla and Quadro,
possibly indicating that CUDA is better optimized to the NVIDIA GPU architectures. 
However, the difference is modest and seems a reasonable price for the flexibility offered by an architecture-independent code.
Comparing Tesla and Quadro, we find the Tesla GPU to be more powerful, which becomes visible at large-enough input data (number of events). 
The hardware differences between these two GPU cards are manifold -- processor speed, global memory size, number of computing cores etc~\cite{c2075, quadro}.
We conclude that the Tesla GPU seems more appropriate for our problem than the Quadro GPU.

\section{Investigations on multi-core CPU}
\label{sec:multicore}

\begin{figure}[tbp]
	\begin{center}
	\includegraphics[width=0.45\linewidth]{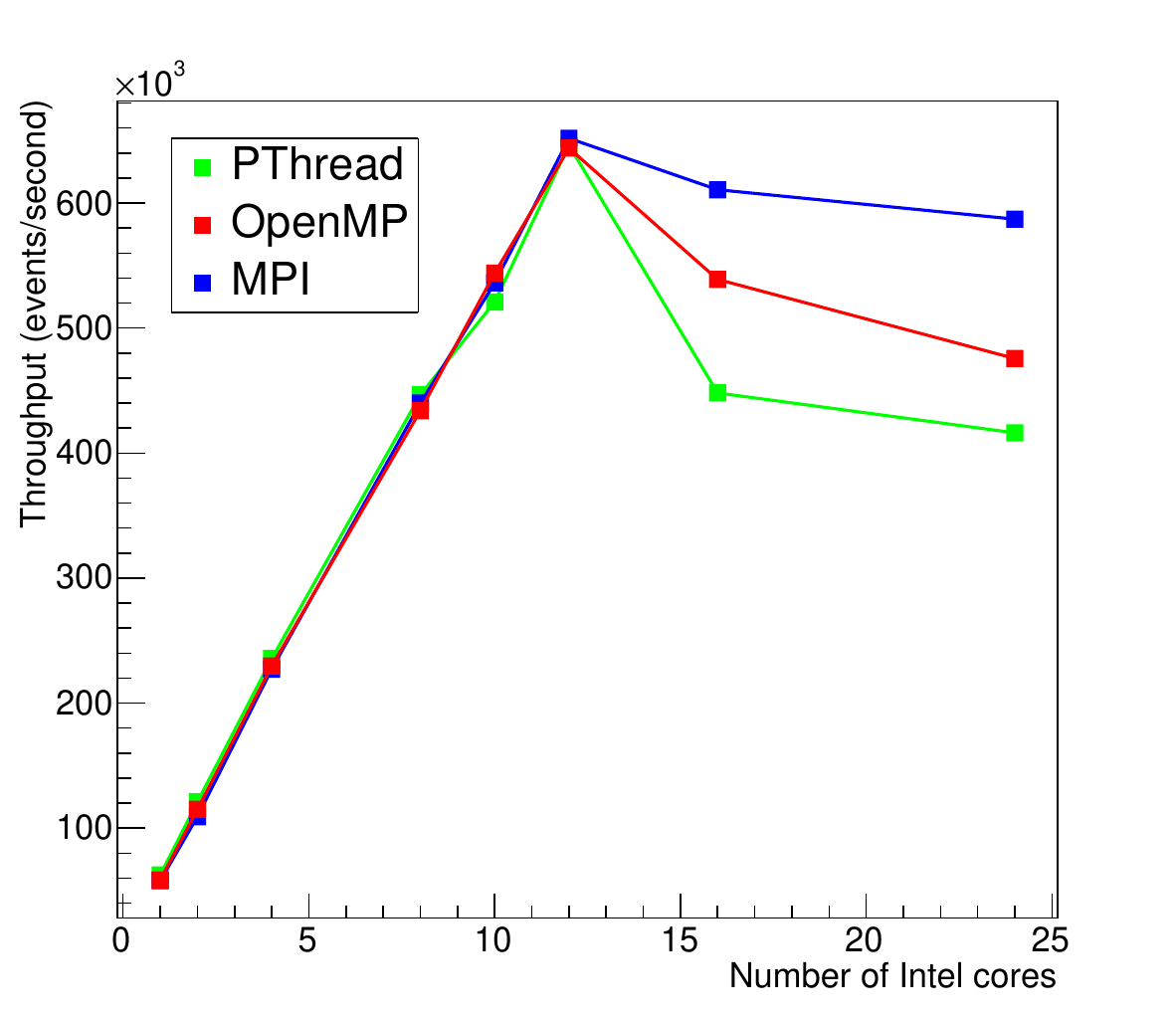}
	\hfill
	\includegraphics[width=0.5\linewidth]{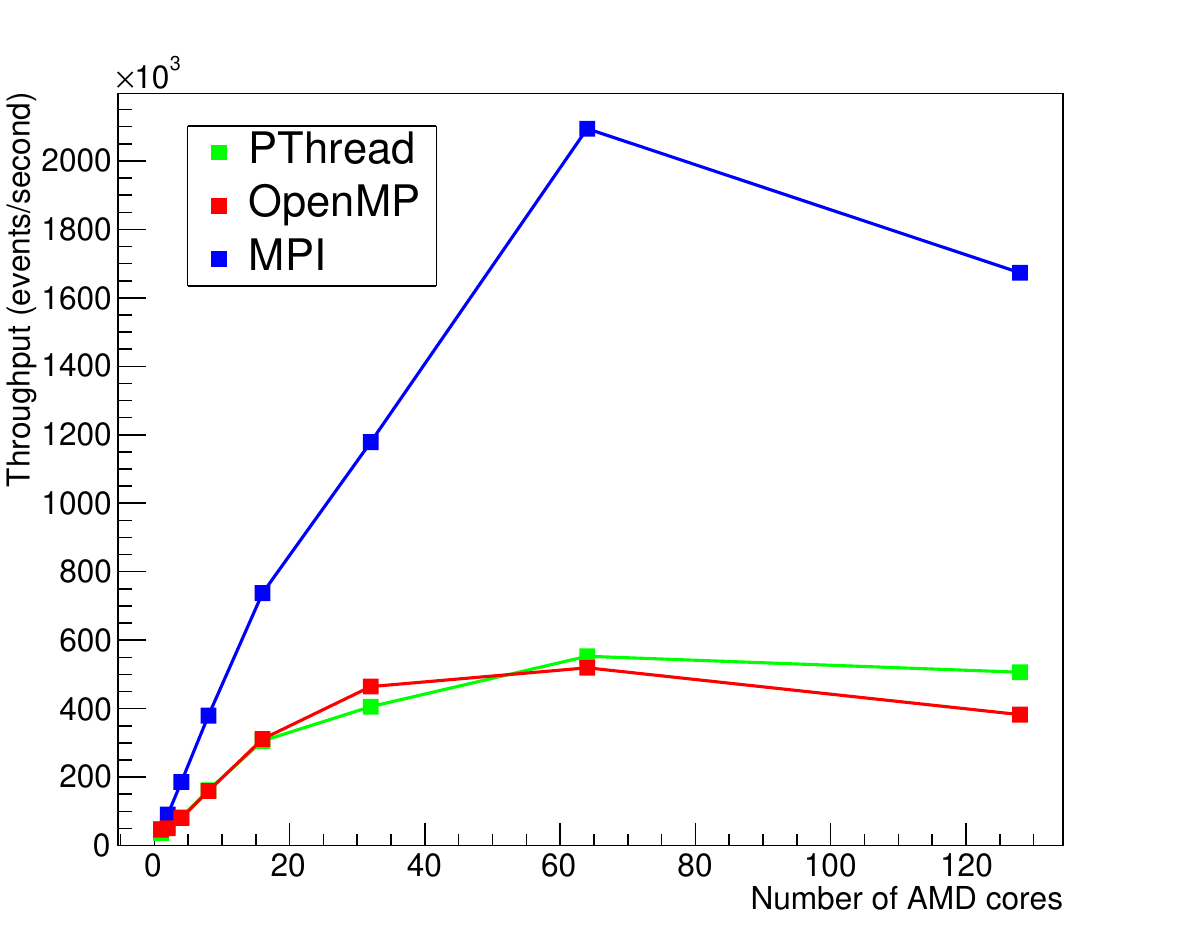}
	\caption{Throughput (number of events executed per second) obtained
	with the pthread, OpenMP and the MPI implementations as a function of the number of cores used in parallel for a sample of 20k events. The left panel shows the results for setup S1 (Intel, 12 physical cores), 
	the right panel those for setup S2 (AMD, 64 physical cores).}
	\label{MPIOpenMP}
	\end{center}
\end{figure}

An alternative to using GPU accelerator co-processors is to make use of the multi-core CPU architecture present in contemporary computers~\cite{HansenBook, QuinnBook}.
Concurrency on CPU cores can be established using pthread, OpenMP, MPI, and OpenCL, all of which are open-source programming paradigms, where OpenCL is primarily developed for many-core or GPU architecture. A preliminary study using pthread and OpenMP only was presented in~\cite{AarushiPaper}, demonstrating the importance of the proper choice of compiler options. 
Here, we study in addition MPI and, in particular, OpenCL. We tested implementations of the event selection algorithm for all four of these programming paradigms on the two platforms S1 and S2; the GPUs of the S1 setup were idle or in open condition. Hardware parallelism was exploited in the simplest way by processing one event per thread (see subsection~\ref{subsec:cuda}). 

Figure~\ref{MPIOpenMP} (left) compares the throughput (number of events executed per second) on the Intel Xeon processors (2 x 6 cores) of setup S1 in dependence of the number of cores (threads) used in parallel for a sample of 20000 events. 
We find for all three pure parallel programming implementations a linear scaling with the number of threads up to 12 threads, from when on the throughput decreases again.
This signals that from this point onwards, the context switching time starts to dominate the total process time.
The same test was performed on the setup S2 (4 x AMD Opteron 16 cores) as shown in the right panel of Figure~\ref{MPIOpenMP}, obtaining similar results for 64 threads. OpenCL treats the underlying device, in this case the CPU, as a single compute unit; therefore, different timing results cannot be gathered by varying the number of cores. 

For both setups, we find the throughput to scale with the number of threads / physical cores (the speed-up is 35 for AMD and 11 for Intel), which is to be expected for pure data-level parallelism.
On the Intel setup, the performance obtained with pthread, OpenMP and MPI are similar, where MPI shows slight higher throughput. On the AMD setup, both pthread and OpenMP are less performant than MPI, although hardware-specific compiler flags were used. 
As was already pointed out in section~\ref{sec:computing}, thread spawning and distributing time are not accounted for the MPI implementation; they contribute in proportion to the number of threads. We attribute our findings to the fact that the event selection process does not use shared memory or inter-thread communication as explained earlier in section~\ref{sec:nvidia}. 
Unlike the other frameworks, MPI statically binds the thread to CPU cores.
The similarity of the results for the pthread and the OpenMP implementations are to be expected since internally, OpenMP uses pthread for spawning multiple threads.

\begin{figure}[tbp]
	\begin{center}
	\includegraphics[width=0.8\linewidth]{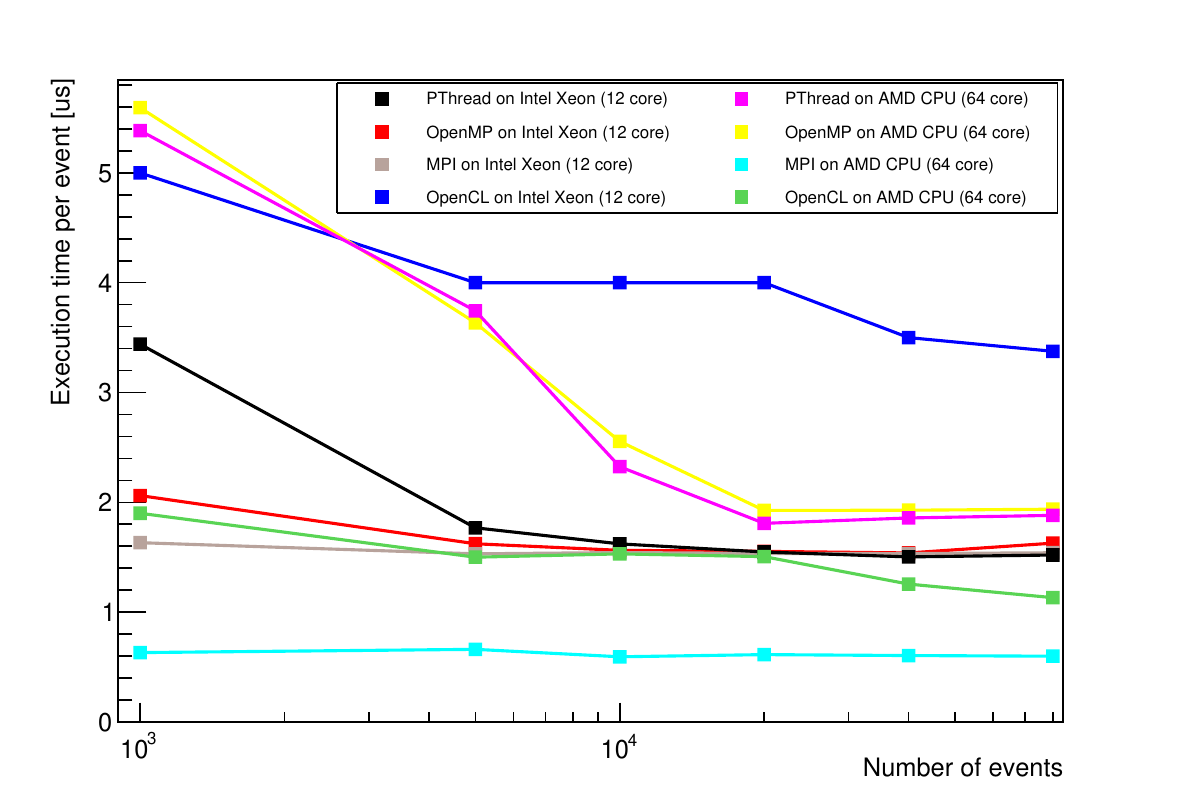}
	\caption{Execution time per event as function of the number of events processed at a time for the implementations with pthread, OpenMP, MPI and OpenCL on the Intel and AMD CPUs. The number of threads equals that of the available physical cores (12 for Intel, 64 for AMD).}
	\label{MPIOpenMPOpenCLCompare}	
	\end{center}
\end{figure}

The performances obtained with OpenCL, pthread, OpenMP and MPI on the two hardware architectures are compared in Fig.~\ref{MPIOpenMPOpenCLCompare} for different numbers of events processed at a time (up to 80,000). 
The number of threads is 12 for Intel and 64 for AMD, as shown to be optimal by Fig.~\ref{MPIOpenMP}. 
On both platforms, we find the execution times for pthread and OpenMP to decrease with the number of events and then saturate,
indicating a minimal data size (about 20k events) from which on the process overhead can be neglected.
On Intel, OpenCL performs clearly worse than the other implementations, whereas it is slightly better than OpenMP and pthread on AMD; here, MPI is found to clearly give the best execution speed.
As reasons for OpenCL to perform worse than e.g., MPI, we have to acknowledge the fact that OpenCL was primarily designed 
as a GPU programming tool; thus, its performance is proportional to the number of thread invocations. 
OpenCL also produces vectorized code in an automatized way, whereas manual vectorization and/or compiler optimization flags need being used for better performance on other implementations. 

A comparison of the Intel and AMD processors is not straightforward because of the differences in the number of computing units and theoretical peak performances.
Considering only the throughput per core, AMD appears less performant than Intel for all implementations since the number of parallel threads is 5 times larger than for the Intel CPUs, but the per event execution time is only 2.5 times smaller. 
A complete assessment, however, would have to also take into account the costs for purchase and operation, which is beyond the scope of this article.

\section{Conclusions}
\label{sec:conclusions}

We have described the development of an event selection algorithm for the CBM-MUCH detector data and a systematic study for the implementation of the event selection process using different parallel computing paradigms like pthread, OpenMP, MPI, and OpenCL for multi-core CPU architectures, and  CUDA and OpenCL for many-core architectures like NVIDIA GPUs. 
For both platforms, the event selection procedure suppresses the archival data rate by almost two orders of magnitude without reducing the signal efficiency, thus satisfying the CBM requirements for high-rate data taking.

On GPUs, we have found a speed-up of 4.5 with respect to the optimized single-thread execution on CPU. This result, however, is only obtained after careful optimization of the implementation in CUDA.
OpenCL on NVIDIA GPUs are found to perform only slightly worse than that for CUDA.
Our results show that about $3 \cdot 10^5$ events per second can be processed on a single GPU card of NVIDIA Tesla family.
Present hardware supports up to four GPUs on a single motherboard. 
This suggests that the targeted CBM interaction rate of $10^7$ events per second can be accommodated by a small number of servers properly equipped with GPUs.

In a multi-core CPU environment, we have compared OpenCL, pthread, OpenMP and MPI as open-source concurrency paradigms. 
A linear scaling of the data throughput with the number of parallel threads is observed up to the number of available physical cores.
In the powerful S2 setup with in total 64 AMD cores, we find that about $2 \cdot 10^6$ events can be processed per second, which is already close to the targeted event rate of $10^7$/s. This demonstrates that SIMD instructions provided by modern CPUs are essential to achieve the required throughput, and that the computing demands of the CBM experiment for the real-time selection of J/$\psi$ candidate events can be achieved by properly making use of the parallel capacities of heterogeneous computing architectures. As an example, the NVIDIA Tesla GPU of setup S1 could be placed into setup S2 to achieve the desired goal.

Comparing the different programming paradigms, we find the cross-platform OpenCL to be a proper choice for heterogeneous computing environments typical for modern architectures, which combine CPU cores with GPU-like accelerator cards.
For such kind of systems, OpenCL provides a suitable solution to simultaneously exploit all available compute units for a given application.
It also provides the flexibility to future improvements in computing architectures, which is of particular importance for CBM as an experiment in the construction stage.
This flexibility, however, comes at the price of a reduced performance on CPU when compared to pure parallel programming paradigms.

\section{Outlook}
The data selection procedure developed and investigated in this article relies on data aggregated into events, corresponding to a single nucleus-nucleus interaction.
The data acquisition of the CBM experiment, however, will deliver free-streaming data not associated to single event by a hardware trigger.
To properly account for this situation, not only the spatial coordinates, but also the time measurement of each hit must be considered.
This will increase the complexity of the current, rather simple algorithm.

We are working together with the CBM collaboration towards extending the algorithm to event building and selection from the real online data stream and also will investigate the throughput on multi-core and many-core platforms in parallel using hybrid programming~\cite{HybridOpenmpMpi}. In addition, other algorithmic approaches to the trigger problem will be investigated, reducing the combinatorics by a more selective triplet construction.

Our study shows that the computational problem can be solved with reasonable expenditure
on CPUs, but also on GPUs as co-processors, or by a combination of both~\cite{MpiCuda, CudaOpenMpMPI}. 
It does not yet include a full exploitation of possible measures for further acceleration, like using vendor-specific compilers (Intel) or using manual code vectorization.
Such investigations will be performed in the future as prerequisites for a decision on the hardware architecture, which of course will have to balance performance with acquisition and running costs.

\section*{Acknowledgments}
{We would like to express our sincere gratitude to the CBM Collaboration for all helps. We are also very grateful to Dr. Partha Pratim Bhaduri of VECC Kolkata for his valuable suggestions and various theoretical concepts during our work. We acknowledge the services and computing facility provided by the grid computing facility at VECC-Kolkata, India. }

\end{document}